\documentclass[prl,aps,amssymb,twocolumn]{revtex4}
\usepackage[dvips]{epsfig,color}
\usepackage{graphicx}
\usepackage{amsmath,amsfonts,amsbsy,amssymb}

\newcommand{\nn}{\nonumber}
\def\tr{\text{tr}}
\def\be{\begin{equation}}
\def\ee{\end{equation}}
\def\bse{\begin{subequations}}
\def\ese{\end{subequations}}
\def\bal{\begin{align}}

\begin{document}

\title{Chiral Topological Insulator on  Nambu 3-Algebraic  Geometry 
}

\author{Kazuki Hasebe}
\affiliation{Department of Physics, Stanford University, Stanford, California 94305, USA \\
\sf
khasebe@stanford.edu\footnote{On leave from Kagawa National College of Technology, 551 Takuma-cho, Mitoyo, Kagawa 769-1192, Japan. \\
 After March 31 2014,  email to \sf{hasebe@dg.kagawa-nct.ac.jp}}}

\begin{abstract}
Chiral topological insulator (AIII-class) with  Landau levels is constructed based on the Nambu 3-algebraic geometry.  We clarify the geometric origin of the chiral symmetry of the AIII-class topological insulator in the context of non-commutative geometry of 4D quantum Hall effect.   
The many-body groundstate wavefunction is explicitly derived as a $(l,l,l-1)$ Laughlin-Halperin type  wavefunction with unique $K$-matrix structure. Fundamental excitation is identified with anyonic string-like object  with  fractional charge ${1}/({2(l-1)^2+1})$.  The Hall effect of the chiral topological insulators turns out be a color version of Hall effect, which exhibits a dual property of the Hall and spin-Hall effects.    

\end{abstract}

\maketitle


\section{Introduction}

In the past decade,  the topological insulars (TIs) with time-reversal symmetry have attracted great attentions. 
The chiral TI is a new class of TI  that has not been experimentally observed. 
The chiral IT is also known as AIII-class TI which  
respects the chiral symmetry and lives in arbitrary odd dimension 
\cite{RyuSFL2009}. 
Since the chiral TI can live in 3D space, the chiral TI is expected to be realized in daily experiments.  
Indeed, the lattice model of the chiral TI has been discussed in Refs.\cite{Hosuretal2009,NeupertSRChM2012}, and its possible 
experimental platform has been proposed in Ref.\cite{WngDD2014}.   


Recently, two groups independently applied  non-commutative geometry (NCG) techniques to 
TIs \cite{NeupertSRChM2012,EstienneRB2012} 
and discussed the appearance of quantum Nambu geometry \cite{Nambu1973,CurtrightZachos2003,DeBellisSS2010}  in the context of TIs. 
Since quantum Nambu geometry is  closely related to the geometry of M-theory \cite{BasuHarvey2004,BaggerLambert2006,Gustavsson2007,BaggerLambert2007}, 
the appearance of quantum Nambu geometry in TIs is quite intriguing, however   
 the two groups reached a contradictory conclusion about the Nambu 3-bracket description for  TIs; The authors of Ref.\cite{NeupertSRChM2012}  
insist that 3-algebra consistently describes physics of the chiral TI, while the authors of 
Ref.\cite{EstienneRB2012} advocated the 3-algebra is not appropriate because of ``pathological''  properties of the 3-algebra. Here arises a question: (i) Which statement is correct or is there any compromise between these two?   
In Ref.\cite{NeupertSRChM2012}, the projection density operator method was applied to derive excitation energy within the single mode approximation, however the calculation cannot completely be carried out due to the lack of  knowledge of the explicit form of the groundstate. 
Then arises the second equation: (ii) How can we reasonably construct the explicit groundstate wavefunction of the chiral TI?   

In Ref.\cite{Hasebe2014-1}, the author clarified  relations between the A-class TIs and  quantum Hall effect (QHE) in arbitrary even dimension. 
A-class and AIII-class TIs share many similar properties: 
Both A-class and AIII-class are classified by $\mathbb{Z}$ topological invariant and regularly appear in even and odd dimensions, and either of them does not  respect  time-reversal or particle-hole symmetry. However there is one discrepancy: AIII-class respects the chiral symmetry  while A-class does not.  
Since A-class TIs  are realized as QHE in even dimensions, 
the AIII-class TIs may be regarded as odd dimensional analogue of QHE.  If so, it is reasonably understood  why A and AIII-class TIs are so much like. At the same time, the third question  arises: (iii) Why does only AIII-class have the chiral symmetry?  There are not many works about QHE in odd dimension except for the pioneering  work of  Nair and Randjbar-Daemi  \cite{Nair-Daemi-2004} where they found the Landau level spectrum depends on a ``mysterious'' extra parameter  whose  couterpart does not exist in the even dimensional case.  
Here arises the last question: (iv)  What is the physical meaning of the extra parameter found in Nair and Randjbar-Daemi's analysis?   

In this paper, we explore 3D chiral TI with emphasis on its relation to quantum Nambu geometry.  Through the work, we provide convincing resolutions to all of  the controversial issues from (i) to (iv). 


\section{The Landau Problem on $S^3$}

We first revisit the $SO(4)$ Landau model on a three-sphere in the $SU(2)$ monopole background \cite{Nair-Daemi-2004}: 
\begin{equation}
H=\frac{1}{2Mr^2}\sum_{\mu <\nu =1}^4{\Lambda_{\mu\nu}}^2, 
\label{hamitonianso4}
\end{equation}
where $r$ denotes the radius of three-sphere. 
The covariant angular momentum is constructed as 
$\Lambda_{\mu\nu}=-ix_{\mu}D_{\nu}+ix_{\nu}D_{\mu}$   
where the covariant derivative is given by  
$D_{\mu}=\partial_{\mu}+iA_{\mu}$   
with $SU(2)$ monopole gauge field 
\be
A_{\mu}=(A_i, A_4)=   
(-\frac{1}{2r(r+x_4)}\epsilon_{ijk}x_j\sigma_k, 0). 
\label{su2gaugefield}
\ee  
Here, $\frac{1}{2}\sigma_i$ denote the $SU(2)$ matrices with spin magnitude $I/2$.  
The corresponding field strength  
$F_{\mu\nu}
=\partial_{\mu}A_{\nu}- \partial_{\nu}A_{\mu}+ i[A_{\mu},A_{\nu}]$  
is given by 
$F_{ij}=-\frac{1}{r^2}x_{i} A_{j}+\frac{1}{r^2}x_{j}A_{i}+\frac{1}{2r^2}\epsilon_{ijk}\sigma_k,$ and $F_{i4}=\frac{1}{r^2}(r+x_4)A_i=-\frac{1}{2r^3}\epsilon_{ijk}x_j\sigma_k$, which satisfy 
$\sum_{\mu<\nu}{F_{\mu\nu}}^2=\frac{1}{4r^4}{\sigma_i}^2=\frac{1}{4r^4}I(I+2)$ and 
$\sum_{\mu<\nu}F_{\mu\nu}\tilde{F}_{\mu\nu}=0,$  
with 
$\tilde{F}_{\mu\nu}=\frac{1}{2}\epsilon_{\mu\nu\rho\sigma}F_{\rho\sigma}.$ 
The Hamiltonian (\ref{hamitonianso4}) may respect the $SO(4)$ symmetry since the the $SU(2)$ monopole magnetic field  is perpendicular to the surface of $S^3$  $(x_{\mu}F_{\mu\nu}=F_{\mu\nu}x_{\nu}=0)$.   
 The $SO(4)$ total  angular momentum is constructed as   
$L_{\mu\nu}=\Lambda_{\mu\nu}+r^2F_{\mu\nu}$,     
which satisfy 
$[L_{\mu\nu},O_{\rho\sigma}]=i\delta_{\mu\rho}O_{\nu\sigma}-i\delta_{\mu\sigma}O_{\nu\rho}+i\delta_{\nu\sigma}O_{\mu\rho}-i\delta_{\nu\rho}O_{\mu\sigma},$  
where $O_{\mu\nu}=L_{\mu\nu}, \Lambda_{\mu\nu}, F_{\mu\nu}$.  One may confirm that $H$ is indeed invariant under the $SO(4)$ transformations, $[H, L_{\mu\nu}]=0$.  
The $SO(4)$ algebra consists of two independent $SU(2)$ algebras, $SU(2)_L\oplus SU(2)_R$, as    
$L_i^{\pm}=
\Lambda_i^{\pm}+F^{\pm}_i,$  
where 
$O_i^{\pm}\equiv \frac{1}{2}\sum_{\mu<\nu}{\eta_{\mu\nu}^{\pm}}^i O_{\mu\nu},$  
with $O=\Lambda, F, L$ and t'Hooft tensor ${\eta_{\mu\nu}^{\pm}}^i=\epsilon_{\mu\nu i 4}\pm \delta_{\mu i}\delta_{\nu 4}\mp\delta_{\mu 4}\delta_{\nu i}$.  
The  $SO(4)$ Landau Hamiltonian can be decomposed to two $SU(2)$ invariant Hamiltonians:   
\be
H=H_L+H_R,  
\ee
where 
$H_{L}=\frac{1}{Mr^2}{\Lambda_i^{+}}^2$ and $H_{R}=\frac{1}{Mr^2}{\Lambda_i^{-}}^2$.  
Due to the relations  
$F_i^{\pm}\Lambda_i^{\pm}=\Lambda_i^{\pm}F_i^{\pm}=0$, they can be rewritten as  
$H_{L/R}=\frac{1}{Mr^2}({L_i^{\pm}}^2-r^4{F_i^{\pm}}^2)= \frac{1}{Mr^2}({L_i^{\pm}}^2-\frac{1}{16}I(I+2)),$  
with 
$\Lambda_i^{\pm}
=\mathcal{L}_i^{\pm}+\frac{1}{4r}\biggl((r-x_4)\delta_{ik}-\frac{1}{r+x_4}x_ix_k\pm\frac{x_4}{r+x_4}\epsilon_{ijk}x_j\biggr)\sigma_k,$  
$\mathcal{L}_i^{\pm}=-i\sum_{\mu<\nu} {\eta_{\mu\nu}^{\pm}}^i x_{\mu}\partial_{\nu}=-i\frac{1}{2}\epsilon_{ijk}x_j\partial_k\mp i\frac{1}{2}x_i\partial_4\pm i\frac{1}{2}x_4\partial_i,$ 
$F_i^{\pm}=\frac{1}{4r^3}(x_4\delta_{ik}+\frac{1}{r+x_4}x_ix_k\pm \epsilon_{ijk}x_j)\sigma_k$,  
and so 
$L_i^{\pm}=\Lambda_i^{\pm}+F_i^{\pm}=\mathcal{L}_i^{\pm}+\frac{1}{4}(\delta_{ik}\mp\frac{1}{r+x_4}\epsilon_{ijk}x_j)\sigma_k.$   
$L_i^+$ and ${L}_i^-$ are interchanged under the ``parity'' transformation: 
$(x_i, x_4)~\rightarrow~(-x_i, x_4)$.  
Notice  that  though $H^{L/R}$ take a superficially similar form of the $SO(3)$ Landau Hamiltonian 
 on the Haldane's sphere \cite{Haldane1983}, they are $SU(2)$ matrix valued Hamiltonians.   
The eigenvalues 
are readily derived as   
$E_{l_L, l_R}=\frac{1}{Mr^2}\biggl(l_L(l_L+1)+l_R(l_R+1)\biggr)-\frac{1}{8Mr^2}I(I+2)$ 
where $l_L$ and $l_R$ denote the $SU(2)_L\otimes SU(2)_R$ angular momentum indices. 
The diagonal $SU(2)_D$ operators are constructed as 
$L_i^D=L_i^++{L}_i^-=-i\epsilon_{ijk}x_j\partial_k+\frac{1}{2}\sigma_i,$  
which obviously satisfy $[L_i^D, L_j^D]=i\epsilon_{ijk}L_k^D$.  

 With a given monopole charge $I/2$, the eigenvalues of the $SU(2)_L$ and $SU(2)_R$ angular momentum indices are related as \cite{Nair-Daemi-2004}
\be
l_L+l_R={n}+\frac{I}{2}, ~~~~~
l_L-l_R=s. 
\label{relatlpmnIs}
\ee
Here $n$ denotes the Landau level index $(n=0,1,2,\cdots)$, and $s$ corresponds to the extra parameter that takes  integer of half-integer values\footnote{We adopt ${I}/{2}$ and $n$ instead of $J$ and $q$ in Ref.\cite{Nair-Daemi-2004}. $s$  is related to the extra parameter $\mu$ in Ref.\cite{Nair-Daemi-2004} by $s=\frac{1}{2}I-\mu$.}.  Therefore,  
$l_L$ and $l_R$ can respectively be expressed as  
$l_{L}=\frac{1}{2}(n+\frac{I}{2}+ s)$ and $l_{R}=\frac{1}{2}(n+\frac{I}{2}-s)$.  
Notice that under  the sign change of  $s$, $L_+$ and $L_-$ are interchanged, and hence $s$ can be identified with the chirality index.  
The energy eigenvalues are rewritten as 
\be
E_n^{(s)}=\frac{1}{2Mr^2}(n(n+2)+\frac{I}{2}(2n+1)+s^2), 
\label{energyeigennands} 
\ee
and the corresponding $n$th Landau level degeneracy is given by 
\begin{equation}
d_n^{(s)}=(2l_L+1)(2l_R+1)=(n+\frac{I}{2}+s+1)(n+\frac{I}{2}-s+1). 
\label{degeracysexplicit}
\end{equation}
In the thermodynamic limit $I, r\rightarrow \infty$ with fixed 
$B=I/(2r^2)$ and finite $s$, $E_n^{(s)}$ reproduces the ordinary Landau level on 2D-plane, 
$\frac{B}{M} (n+\frac{1}{2}).$    
Notice that both of the energy eigenvalue (\ref{energyeigennands}) and the degeneracy (\ref{degeracysexplicit}) depend on  $s$, and  exhibit the chiral symmetry with respect to $s\rightarrow -s$ \footnote{The original definition of the chiral symmetry is  $S \mathcal{H}S^{-1}=-\mathcal{H}$ \cite{RyuSFL2009}. 
This relation indeed holds for the square root of the Hamiltonian (\ref{hamitonianso4})  (Dirac Hamiltonian) with $S$ (\ref{defofq5}). 
}.   
In the lowest Landau level (LLL) $n=0$, the energy is represented as 
\be
E_{\text{LLL}}^{(s)}=\frac{1}{2Mr^2}s^2+\frac{I}{4Mr^2}, 
\label{LLLeigenvaluespara}
\ee
where due to the constraints (\ref{relatlpmnIs}), $s$ takes 
$0, \pm 1,\pm 2, \cdots, \pm(\frac{I}{2}-1), \pm\frac{I}{2}$  for even $I$, while 
${{\pm \frac{1}{2}}}, \pm \frac{3}{2}, \pm \frac{5}{2}, \cdots, \pm (\frac{I}{2}-1), \pm \frac{I}{2}$ {for} odd $I$.  Therefore,  the minimum energy of (\ref{LLLeigenvaluespara}) is achieved at $s=0$ for even $I$, and at $s=\pm 1/2$ for odd $I$. It should be emphasized that for odd $I$, the LLL  has ``two fold'' degeneracy coming from $s=1/2$ and $s=-1/2$: 
\be
D(I)=d_{LLL}^{(s=1/2)}+d_{LLL}^{(s=-1/2)}=\frac{1}{2}(I+1)(I+3).  
\label{totalnumberstatesdI}
\ee


\section{The Chiral Hopf map and quaternions}

Let us consider the LLL basis states for $s=\pm 1/2$. 
We derive their functional form instead of the  abstract Wigner $\mathcal{D}$-function \cite{Nair-Daemi-2004}.   
For this purpose, we first introduce the chiral Hopf map: 
\be
S^3_L\otimes S^3_R~\overset{S^3_D}{\longrightarrow}~ S^3. 
\label{mapfromtwo3spheretoone}
\ee
The coordinates of $S_L^3\otimes S_R^3$ are expressed by  
the two-component complex spinors  $\psi_L$ and $\psi_R$ (chiral Hopf spinors) subject to the normalization condition,  
${\psi_L}^{\dagger}\psi_L={\psi_R}^{\dagger}\psi_R=\frac{1}{2}$,   
and the chiral Hopf map  is explicitly realized as 
\be
\psi_L, ~\psi_R ~\longrightarrow ~\frac{x_{\mu}}{r}={\psi_R}^{\dagger}{q_{\mu}}\psi_L + {\psi_L}^{\dagger}\bar{q}_{\mu}\psi_R,  ~~~(\mu=1,2,3,4)
\label{1.5Hopfmap}
\ee
where $q_{\mu}$ and $\bar{q}_{\mu}$ denote the  quaternions and conjugate-quaternions,    
$q_{\mu}=(q_i, 1)=(-i\sigma_i, 1)$ and 
$\bar{q}_{\mu}=(-q_i, 1 )=(i\sigma_i, 1)$ 
\footnote{$~\bar{}~$ represents the quaternionic conjugation.}. 
It is straightforward to show that $x_{\mu}$ (\ref{1.5Hopfmap}) obey 
$\sum_{\mu=1}^4 x_{\mu}x_{\mu}=4r^2({\psi_L}^{\dagger}\psi_L)\cdot({\psi_R}^{\dagger}\psi_R)=r^2.$  
$x_{\mu}$ are invariant under the simultaneous $SU(2)_D$ transformation of $\psi_L$ and $\psi_R$: 
$\psi_{L/R} \rightarrow \psi_{L/R} e^{\alpha_i q_i}.$  
The explicit form of the chiral Hopf spinors is given by   
\be
\psi_{L/R}(x)=\frac{1}{\sqrt{2}}M_{L/R}(x)\phi, 
\label{su2su2bispinors}
\ee
where $\phi$ denotes a normalized two-component spinor ($S^3$-fibre) with the normalization  $\phi^{\dagger}\phi=1$ and 
$M_{L/R}$ is the following unitary matrix:  
\be
M_{L/R}(x)=\frac{1}{\sqrt{2r(r+x_4)}}(r+ x_{\mu}{q}_{\mu}^{R/L}), 
\label{defofmatrixM}
\ee
with $q^L_{\mu}\equiv q_{\mu}$ and $q^{R}_{\mu}\equiv \bar{q}_{\mu}$, and  ${M}_{R}$  is a  quaternionic conjugate of  $M_L$: 
${M}_R(x)=(M_L(x))^{\dagger}=(M_L(x))^{-1}$.   
$M_{L/R}$ denotes a square root of the $SU(2)$ element $g=\frac{x_{\mu}}{r}\bar{q}_{\mu}$: 
${M_L}^2=g$ and ${M_R}^2=g^{\dagger}$.  
Notice that $\psi_L$ and $\psi_R$ are related by the ``parity transformation'': 
$\psi_R(x_i,x_4)=\psi_L(-x_i,x_4)$,   
which is equivalent to the quaternionic conjugate in (\ref{defofmatrixM}). 
We can derive the $SU(2)$ gauge field (\ref{su2gaugefield}) as $A=-i{M_{R}}dM_L-iM_L dM_R$.  
One may readily verify that $M_{L/R}$ satisfies 
$L_i^{+} M_{L}=\frac{1}{2}M_{L}\sigma_i$,  $L_i^{-} M_{R}=\frac{1}{2}M_{R}\sigma_i$, 
${L}_i^{+}M_{R}={L}_i^{-}M_{L}=0,$  and so  $\psi_L$ and $\psi_R$ respectively transform as  $SU(2)_L\otimes SU(2)_R$ Weyl spinors, $(1/2, 0)$ and $(0, 1/2)$. The direct product of the two Weyl spinors gives the $SU(2)_L\otimes SU(2)_R$ ``bi-spin'' representation $(l_L, l_R)$   that 
 corresponds to the LLL basis state for $(l_L, l_R)=(\frac{1}{2}(\frac{I}{2}+s), \frac{1}{2}(\frac{I}{2}-s))$. 
The $(l_L, l_R)$  representation is explicitly constructed as 
\be
\Psi^L_{l_L, m_L}\otimes \Psi^R_{l_R, m_R}, \
\ee
where 
\be
\Psi_{l_L, m_L}^{L}=\frac{1}{\sqrt{(l_L+m_L)!(l_L-m_L)!}}(\psi^1_{L})^{l_L+m_L}(\psi^2_{L})^{l_L-m_L}, 
\ee
with 
$m_L=-l_L, -l_L+1, \cdots , l_L-1, l_L$. 
Same for $\Psi^{R}_{l_R,m_R}$ by replacing $L$ with $R$. 
Thus, the LLL basis states are given by the holomorphic function of $\psi_L$ and $\psi_R$.  
The chiral Hopf map is naturally derived by the ``dimensional reduction'' of the 2nd Hopf map, $S^7\overset{S^3}\rightarrow S^4$.  The 2nd Hopf map  is realized as a map from a four-component complex spinor $\psi$ subject to $\psi^{\dagger}\psi=1$ to  $x_a=\psi^{\dagger}\gamma_a\psi$  $(a=1,\cdots, 5)$ with 
the $SO(5)$ gamma matrices, 
$\{\gamma_{\mu}, \gamma_5\}=\{\begin{pmatrix}
0 & \bar{q}_{\mu} \\
{q}_{\mu} & 0 
\end{pmatrix}, \begin{pmatrix}
1_2 & 0 \\
0 & -1_2
\end{pmatrix}\}$ \cite{ZhangHu2001},  
 and the chiral Hopf map is obtained by imposing  an additional constraint $\psi^{\dagger}\gamma_5 \psi={\psi_L}^{\dagger}\psi_L -{\psi_R}^{\dagger}\psi_R=
0$ with $\psi=(\psi_L, \psi_R)^t$.  
This   
implies a geometric embedding of the chiral TI in 4D QHE. 
Similarities between the chiral TI and 4D QHE can also be found in  the $SU(2)$-bundle topology.  
With use of the chiral Hopf spinors, 
$Q$ matrix in Ref.\cite{RyuSFL2009} is derived  as 
$Q=1-\begin{pmatrix}
M_L \\
M_R
\end{pmatrix} ({M_R}~ {M_L})
=-\begin{pmatrix}
0 & g \\
g^{\dagger} & 0 
\end{pmatrix},$  
and the corresponding winding number is evaluated as  
$c_2 =-i\frac{1}{24\pi^2} \int_{S^{3}} \tr (-ig^{\dagger}dg )^{3} 
=\frac{1}{6}I(I+1)(I+2),$  
which is exactly equivalent to the 2nd Chern number of the $SU(2)$ monopole-bundle over $S^{4}$ -- the set-up of 4D QHE. 

\section{Quantum Nambu Geometry}

Since the LLL basis states are given by the holomorphic function of the chiral Hopf spinors,  the complex conjugation can be regarded as the derivative, ${\psi_{L, R}}^*\rightarrow {\partial}/{\partial \psi_{L, R}}$. 
From the chiral Hopf map, we obtain the effective operator expression for the $S^3$ coordinates:   
\be
X_{\mu}
=\frac{\alpha}{2} {\psi_R}^t q_{\mu}\frac{\partial}{\partial \psi_L}+\frac{\alpha}{2} {\psi_L}^t \bar{q}_{\mu}\frac{\partial}{\partial \psi_R}, 
\label{xmufuzzy3sphere} 
\ee
where $\alpha=2r/I$. 
From the algebras of quaternions\footnote{  $q_{\mu}\bar{q}_{\nu}-q_{\nu}\bar{q}_{\mu}=-2{{\eta}_{\mu\nu}^-}^i q_i$ and  $\bar{q}_{\mu}{q}_{\nu}-\bar{q}_{\nu}q_{\mu}=-2{\eta_{\mu\nu}^+}^i q_i$},   we have 
\be
[{X}_{\mu}, {X}_{\nu}]=i\alpha({\eta_{\mu\nu}^+}^i {X}_i^++{{\eta}_{\mu\nu}^-}^i {X}_i^-), 
\label{chiral4dalgebra} 
\ee
where ${X}^{+}_i=\frac{\alpha}{2}{\psi_{L}}^t {\sigma}_i \frac{\partial}{\partial \psi_{L}}$  and 
${X}^{-}_i=\frac{\alpha}{2}{\psi_{R}}^t {\sigma}_i \frac{\partial}{\partial \psi_{R}}$ are two independent $SU(2)$ operators.  
 Eq.(\ref{chiral4dalgebra}) realizes the chiral symmetric version of the NC algebra of 
4D QHE \cite{ZhangHu2001}.  
We also have 
$[X_{\mu}, X_i^\pm]=-i\frac{\alpha}{2}{\eta_{\mu\nu}^\pm}^i X_{\nu}^\pm$.   
In total, the ten operators, $X_{\mu}$, $X_i^+$ and $X_i^-$, amount to form  the $SO(5)$ algebra. 
The parameter $s$ (\ref{relatlpmnIs}) denotes the eigenvalue of the chiral charge operator: 
\be
S\equiv \frac{1}{2}{\psi_L}^t \frac{\partial}{\partial \psi_L}- \frac{1}{2}{\psi_R}^t \frac{\partial}{\partial \psi_R}.   
\label{defofq5}
\ee
Meanwhile in the set-up of 4D QHE, the 5th coordinate of fuzzy four-sphere is  given by 
$X_5=\frac{\alpha}{2}\psi^t \gamma_5 \frac{\partial}{\partial \psi}= \frac{\alpha}{2}({\psi_L}^t  \frac{\partial}{\partial {\psi_L}}-{\psi_R}^t\frac{\partial}{\partial {\psi_R}})$, and then  
$S=\frac{1}{\alpha}X_5$.   
Remember in the set-up of 3D QHE, $s$ was just an internal parameter, but in the ``virtual'' 4D  QHE,  $s$ can be interpreted as the latitude of $S^4$.   
The chiral symmetry is realized as the reflection symmetry of the 4D QHE with respect to the equator. This is the resolution for (iv). 


In the precedent studies of NCG \cite{Ramgoolam2002,JabbariTorabian2005}, more elegant formulation of the fuzzy three-sphere based on the quantum Nambu bracket has been known  \cite{BasuHarvey2004,JabbariTorabian2005, DeBellisSS2010}. One may readily confirm that matrix realization of $X_{\mu}$ (\ref{xmufuzzy3sphere}) indeed satisfy the quantum Nambu-algebra for the fuzzy three-sphere:    
\be
[X_{\mu},X_{\nu},X_{\rho}]_{\chi}=(I+2)\alpha^2\epsilon_{\mu\nu\rho\sigma}X_{\sigma},  
\label{chiralbracketfuzzys3}
\ee
where the chiral 3-bracket is defined as 
$[X_{\mu},X_{\nu},X_{\rho}]_{\chi} \equiv X_{[\mu}X_{\nu}X_{\rho]}S-X_{[\nu}X_{\rho} S X_{\mu]}+X_{[\rho} S X_{\mu}X_{\nu]}-SX_{[\mu}X_{\nu}X_{\rho]}.$  
Under the ordinary definition of the quantum Nambu 3-bracket, $[X_{\mu},X_{\nu}, X_{\rho}]\equiv X_{[\mu}X_{\nu}X_{\rho]}$, $X_{\mu}$ (\ref{xmufuzzy3sphere}) do not form a closed algebra.  
Here, several comments are added. 
Firstly,  the chiral 3-bracket 
can evade the pathological property of the 3-bracket emphasized in Ref.\cite{EstienneRB2012} as found $[X_{\mu}, X_{\nu}, 1]_{\chi}=0$. Though  the ordinary definition of the 3-bracket was adopted in Ref.\cite{NeupertSRChM2012}, the whole 3-bracket algebra was not really used in the analysis, 
and hence the pathological property of the 3-bracket did not apparently appear.  This gives  the  resolution for (i).     
Secondly, the chiral 3-bracket is neatly fitted in the four-bracket of 4D QHE \cite{Hasebe2014-1} and concisely given by  $[X_{\mu}, X_{\nu},  X_{\rho}]_{\chi}=\frac{1}{\alpha}[X_{\mu}, X_{\nu}, X_{\rho}, X_5]$.   
This is the algebraic evidence that the chiral TI naturally realizes as a ``subspace'' embedded in  the 4D QHE.  Thirdly, the right-hand side of (\ref{chiralbracketfuzzys3}) suggests the existence of 
 3-rank $U(1)$ magnetic field 
$G_{\mu\nu\rho}=\frac{1}{r^4}\epsilon_{\mu\nu\rho\sigma}x_{\sigma}$. 
The corresponding gauge field is given by a 2-rank antisymmetric tensor field 
$C_{\mu\nu}=-C_{\nu\mu}$ 
($G_{\mu\nu\rho}=\partial_{\mu}C_{\nu\rho}+\partial_{\nu}C_{\rho\mu}+\partial_{\rho}C_{\mu\nu}$)  
that couples to string-like object. Such 2-rank tensor  field is simply obtained by the dimensional reduction of the 3-rank tensor gauge field $C_{abc}$  in the 4D QHE 
\cite{Hasebe2014-1} 
by $C_{\mu\nu}\equiv C_{\mu\nu 5}$, and so 
the string-like object 
from  membrane-like excitation in 4D QHE.  

\begin{figure}[tbph]\center
\includegraphics*[width=90mm]{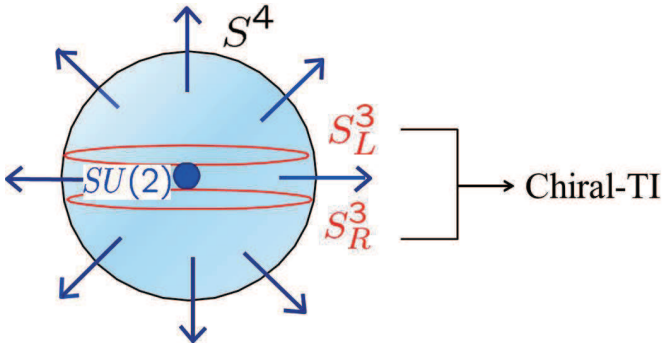}
\caption{Chiral TI is a superposition of  two three-spheres embedded in the 4D QHE, which realizes the chiral symmetry in a geometrical way.    }
\label{AIIIS3}
\end{figure}

The chiral Nambu 3-algebra gives a crucial implication for the existence of the chiral symmetry. The precedent studies  \cite{Ramgoolam2002, JabbariTorabian2005} tell that the fuzzy three-sphere is realized as  a composite of two latitudes $s=1/2$ and $s=-1/2$ of fuzzy four-sphere not just as the  equator ($s=0$). The reason is  simple; If the fuzzy three-sphere was simply  the equator, the NC algebra would  vanish,  $[X_{\mu},X_{\nu}, X_{\rho}]_{\chi}= [X_{\mu},X_{\nu}, X_{\rho}, S]=0$ ($s=0$).  To incorporate a non-trivial NC structure, the fuzzy three-sphere has to be a  composite of two  $S^3$s with  opposite 
 latitudes of same magnitude \cite{JabbariTorabian2005}. This suggests that, 
in the language of TIs,  the chiral TI is given by a superposition of  two $S^3$s  with opposite chiral charges  on the virtual 4D QHE [Fig.\ref{AIIIS3}]. In other words, the requirement of  NCG $\it{necessarily}$ induces the chiral symmetry to the chiral TI. 
  This is the resolution for (iii).  $s=\pm1/2$ is achieved as the minimum energy of the Landau Hamiltonian for odd $I$ not even $I$, and so 
the system becomes  trivial or non-trivial depending on the parity of $I$. This implies a hidden $\mathbb{Z}_2$ structure of the chiral TI. 





\section{Many-body Physics}

 Regarding the two $s=\pm 1/2$ latitudes as ``spin'' degrees of freedom, we apply the Halperin's arguments \cite{Halperin1983}  to explore  many-body physics of the chiral TI.  
The LLL basis states on the $s=+1/2$ latitude, $S^3_L$,  are given by  
$\Psi_{M_L}\equiv \Psi^L_{l_L,m_L}\otimes \Psi_{l_R,m_R}^R|_{(l_L,l_R=\frac{I+1}{4},\frac{I-1}{4})}$, 
and those on the $s=-1/2$ latitude, $S_R^3$, are 
$\Psi_{ M_R}\equiv \Psi^L_{l_L, m_L}\otimes \Psi_{l_R,m_R}^R|_{(l_L,l_R=\frac{I-1}{4},\frac{I+1}{4})}$.   
The total degeneracy  is   $D(I)$ (\ref{totalnumberstatesdI}). 
The Slater determinants on $S_L^3$ and $S_R^3$ are respectively  constructed as 
$\Psi_{\text{L-Slat}}= \epsilon_{M_{{L}_1} M_{{L}_2} \cdots M_{{L}_{N/2}}}\Psi_{M_{L_1}}(x_1)\Psi_{M_{L_2}}(x_2)\cdots \Psi_{M_{L_{N/2}}}(x_{N/2})$, 
$\Psi_{\text{R-Slat}}= \epsilon_{M_{{R}_1} M_{{R}_2} \cdots M_{{R}_{{N/2}}}}\Psi_{M_{R_1}}(x_{N/2+1})\cdots \Psi_{M_{R_{N/2}}}(x_N)$,  
and the Laughlin-Halperin type groundstate wavefunction is given by   
$\Psi_{l,l,m}\equiv \Psi_{\text{L-Slat}}^l\cdot \Psi_{\text{R-Slat}}^l \cdot {\Psi}_{\text{Corr}}^m$ ($l$: odd). 
Here, ${\Psi}_{\text{Corr}}^m$ denotes the correlation part between  $S^3_L$ and $S^3_R$. We can derive the explicit form of ${\Psi}_{\text{Corr}}^m$  by observing that the ``spin-polarized ''state $(l,l,m)=(l,l,l)$ coincides with the Laughlin wavefunction of the total particles: 
$\Psi_{l,l,l}= \Psi_{\text{Llin}}^{(l)} (\equiv 
\Psi_{\text{T-Slat}}^l),$  
where 
$\Psi_{\text{T-Slat}}= \epsilon_{M_{T_1} M_{T_2} \cdots M_{T_N}} \Psi_{M_{T_1}}(x_1)\Psi_{M_{T_2}}(x_2)\cdots \Psi_{M_{T_N}}(x_N)$   
with  $\Psi_{M_{T}}$ $(M_T=1,2,\cdots,D(I))$ denoting the total basis states:  
$\Psi_{M_T}=\{\Psi_{ M_L}, \Psi_{ M_R}\}.$  
The correlation function is determined as   
$\Psi_{\text{Corr}}={\Psi_{\text{T-Slat}}}/({ \Psi_{\text{L-Slat}}\cdot \Psi_{\text{R-Slat}}}). $ 
Hence we have  
\be
\Psi_{l,l,m}=\Psi_{\text{L-Slat}}^{l-m}\cdot \Psi_{\text{R-Slat}}^{l-m} \cdot {\Psi}_{\text{T-Slat}}^m. 
\label{llmequallrt}
\ee
For $\Psi_{l,l,m}$ to be a chiral symmetric state, $i.e.$ the $SU(2)$ singlet state in terms of chiral rotations, $\Psi_{l,l,m}$ has to satisfy the Fock condition and $m$ is restricted to $m=l-1$ \cite{Yoshiokaetal1988}.  Therefore, the chiral (symmetric) wavefunction is given by  
\be
\Psi_{l,l,l-1}=\Psi_{\text{L-Slat}}\cdot \Psi_{\text{R-Slat}}\cdot {\Psi}_{\text{T-Slat}}^{l-1}. = \frac{\Psi_{\text{Llin}}^{(l)}}{\Psi_{\text{Corr}}}.   
\ee
%
This gives a resolution for (iii). 
Under the scaling $I\rightarrow lI$, the total degeneracy behaves as $D(lI)\sim (lI)^2$ and so $\nu= N/D(lI)\sim 1/l^2$. 
To derive a precise expression of the filling factor for $(l,l,m)$  state (\ref{llmequallrt}), 
 we introduce the $K$-matrix \cite{WenZee1992}:   
\be
K=
\begin{pmatrix}
(l-m)^2+m^2 & m^2   \\
 m^2 &  (l-m)^2+m^2
\end{pmatrix}. 
\ee
The $K$ matrix condition is given by 
$K\begin{pmatrix}
N_L \\
N_R
\end{pmatrix}=D
\begin{pmatrix}
1 \\
1 
\end{pmatrix}.$  
Except for  $m=l$,   $K$ has the inverse and   
the filling factors  are derived as 
$\begin{pmatrix}
\nu_L \\
\nu_R
\end{pmatrix}=\frac{1}{D}
\begin{pmatrix}
N_L \\
N_R
\end{pmatrix}=K^{-1}
\begin{pmatrix}
1 \\
1
\end{pmatrix}$,   
and the total filling factor  is  
\be
\nu=\nu_L+\nu_R=\frac{2}{2m^2+(l-m)^2}. 
\ee
The fractional charges are  given by 
$(e^*_L, e^*_R)=({K^{-1}}_{11}, {K^{-1}}_{21})$ or $(e^*_L, e^*_R)=({K^{-1}}_{12}, {K^{-1}}_{22})$,  
and in either case the net fractional charge reads as    
$e^*=e^*_L+e^*_R=\frac{1}{2}\nu$. 
In particular for the $(l, l, l-1)$ state, we have  
\be
e^*=\frac{1}{2}\nu=\frac{1}{2(l-1)^2+1}=1, ~\frac{1}{9},~\frac{1}{33}, ~\frac{1}{73},~\cdots. 
\ee

The coherent state aligned to the direction $\Omega_{\mu}$ ($\sum_{\mu=1}^4{\Omega_{\mu}}^2=1$) on $S^3$  satisfies the quaternionic coherent state equation: 
\be
\Omega_{\mu}\tilde{q}_{\mu}\chi_L=\chi_R, ~~~~~\Omega_{\mu}\bar{\tilde{q}}_{\mu}\chi_R=\chi_L, 
\ee
where 
$\tilde{q}_{\mu}= (-q_1, q_2, -q_3, 1)=(i\sigma^*_i, 1)$ and   $\bar{\tilde{q}}_{\mu}= (q_1, -q_2, q_3, 1)=(-i\sigma^*_i, 1)$. 
Obviously  $\chi_{L}$ and $\chi_{R}$ give 
\be
{\chi_R}^{\dagger}\tilde{q}_{\mu}\chi_L+{\chi_L}^{\dagger}\bar{\tilde{q}}_{\mu}\chi_R=\Omega_{\mu}, 
\ee
and $\chi_{L}$ and $\chi_R$  are expressed as 
$\chi_L=\frac{1}{2\sqrt{1+\Omega_4)}}
(1+\Omega_{\mu}\bar{\tilde{q}}_{\mu})\phi$  
and 
$\chi_R=\frac{1}{2\sqrt{1+\Omega_4)}}
(1+\Omega_{\mu}\tilde{q}_{\mu})\phi$   
with a normalized two-component spinor $\phi$. 
The  point on $S^3$ in the LLL is denoted as $\Omega_{\mu} X_{\mu}$.  
With use of the property of the quaternion $q_2 \tilde{q}_{\mu}={q}_{\mu}q_2$, 
it is readily shown that 
$\psi_{\chi}^{(I)}=({\chi_L}^{\dagger}\psi_L+{\chi_R}^{\dagger}\psi_R)^I$  
 satisfies the coherent state equation: 
$\Omega_{\mu}X_{\mu} \psi_{\chi}^{(I)}=I\psi_{\chi}^{(I)}.$    
Creation and annihilation operators for the charged excitation generated at the point $\Omega_{\mu}(\chi)$ that satisfy 
$[\Omega_{\mu}X_{\mu}, A^{\dagger}(\chi)]=N A^{\dagger}(\chi)$   
and 
$[A(\chi), A^{\dagger}(\chi)]=1,$   
are  constructed as  
\begin{align}
&A^{\dagger}(\chi)
= \prod_{i=1}^N ({A_L}^{\dagger}(\chi_L)_i+{A_R}^{\dagger}(\chi_R)_i),  \nn\\
&A(\chi)
=
\prod_{i=1}^N (A_L(\chi_L)_i+A_R(\chi_R)_i), 
\label{excitationnonchiral}
\end{align}
where 
${A_{L}}^{\dagger}(\chi_{L})_i\equiv i\chi^{t}_L\sigma_2\psi_{L}(i)$,  $A_{L}(\chi_{L})_i\equiv i{\chi_{L}}^{\dagger} \sigma_2 \frac{\partial}{\partial \psi_{L}(i)}$ and  similar expressions for $R$.  
The chiral operators $A_L(\chi_L)_i$ and $A_R(\chi_R)_i$ satisfy 
$[A_L(\chi_L)_i, {A_L}^{\dagger}(\chi_L)_j]=[A_R(\chi_R)_i, {A_R}^{\dagger}(\chi_R)_j]=\delta_{ij},$ and 
$[A_L(\chi_L)_i, A_L(\chi_L)_j]=[A_L(\chi_L)_i, {A_R}^{\dagger}(\chi_R)_j]=[A_R(\chi_R)_i, A_R(\chi_R)_j]=0.$  
Due to the relation $\chi_L^{\dagger}\chi_L=\chi_R^{\dagger}\chi_R =1/2$, either of ${A_L}^{\dagger}(\chi_L)$ and  ${A_R}^{\dagger}(\chi_R)$ cannot be zero. 
Since $\psi=(\psi_L, \psi_R)^t$  is a $SO(4)$ Dirac spinor that carries the $SU(2)_L\oplus SU(2)_R$ bispin $(j_L, j_R)=(1/2, 0)\oplus (0, 1/2)$, $A(\chi)$ and $A(\chi)^{\dagger}$  denote non-chiral operators for charged excitation  
with left and right chiralities.


\section{The Color Hall Effect}

While the chiral TI shares similar properties with QHE such as  time-reversal breaking and    $\mathbb{Z}$ classification of topological invariant,   $\mathbb{Z}_2$ structure is also incorporated in 
the chiral TI due to the chiral symmetry, just like the time-reversal symmetry of the QSHE.  Therefore, the chiral TI is expected to accommodate a dual property of the QHE and QSHE, and such dual property is manifest in 
the transport phenomena.  

Name the two $SU(2)$ color indices $L$ and $R$ and the three colors of $SU(2)$ gauge field    $a=1,2,3$.  
Since the color gauge fields are independently coupled to the corresponding color currents, the Hall effect is given by 
\be
J_i^a=\sigma\epsilon_{ijk} E_j^a B_k^a. ~~~(\text{no sum for $a$}) 
\label{conpletecolorhalleffect}
\ee
Without loss of generality, we focus on  $a=3$ in which  $J_i^3=J^L_i-J^R_i$. 
If there only exist either of  $L$ or $R$-color particles, the Hall effect will be given by  
$J_i^L=\sigma_L\epsilon_{ijk}E_j B_k$ 
and  
$J_i^R=-\sigma_R\epsilon_{ijk}E_j B_k$. 
The $L$ and $R$-color currents flow in the mutually opposite direction,  
similar to the spin Hall effect where the flows of 
up and down spin currents are opposite. 
However, the color Hall effect does not respect the  time reversal symmetry, since 
 the  $L$ and $R$  
 are just labels and are not flipped under the time reversal transformation  unlike physical spin of the spin Hall effect. The time reversal transformation just reverses the direction of the  $L$ and $R$-color currents as in the case of  the ordinary Hall effect. 
The quantized version of the color Hall effect can similarly be understood.   
$L$ and $R$-color currents respectively contribute to the quantized Hall conductivity as 
$\sigma_L=\frac{e^2}{2\pi \hbar} \nu_L,$ 
$\sigma_R=-\frac{e^2}{2\pi \hbar} \nu_R$. 
The total and different conductances are obtained as 
\begin{align}
&\sigma\equiv \sigma_L+\sigma_R= (\nu_L-\nu_R) \frac{e^2}{2\pi \hbar},\nn\\
&\Delta \sigma\equiv \sigma_L-\sigma_R= (\nu_L+\nu_R) \frac{e^2}{2\pi \hbar}.   
\end{align}
For the $(l, l, l-1)$ chiral TI,  
we have a non-chiral version of the Hall effect: 
$\sigma=0$, 
$\Delta \sigma= \frac{1}{2(l-1)^2+1}\frac{e^2}{\pi \hbar}$,  which   reduces to the     
QSH conductance $\Delta\sigma_{\text{QSH}}=\frac{e^2}{\pi \hbar}$  \cite{BernevigZhang2006} for $l=1$.

\section{summary and discussions}



To summarize, we explored one-particle and many-body physics of the chiral TI with Landau levels based on the Nambu 3-algebraic geometry.  
The chiral TI  is a natural 3D generalization of the Haldane's 2D QHE and 3D ``reduction'' of the Zhang and Hu's 4D QHE.  We elucidated the former controversial problems by exploiting the mathematics and physics of the chiral TI.   
In particular, we clarified that Nambu 3-algebraic geometry is essential for the existence of the chiral symmetry of the chiral TI.  
Interestingly, the chiral TI 
exhibits a dual property of  QHE 
and QSHE 
 due to the hidden $\mathbb{Z}_2$ structure of the chiral symmetry. 

Recently, 3D AII TI model with Landau levels was constructed by Li and Wu \cite{LiWu2013} and  Dirac-type models in higher dimensions were also explored in Ref.\cite{Li-I-Y-W-2012}. 
Though the Li and Wu's model also heavily utilized quaternionic structure, the model respects the time-reversal symmetry and hence describes different physics.





\section*{ACKNOWLEDGEMENTS}
\vspace{0.3cm}
 I am very grateful to Yingfei Gu, Pavan Hosur, Biao Lian, V. P. Nair, Xiaoliang Qi, Jing Wang, Shoucheng Zhang, and Yi Zhang for useful discussions. 
I also would like thank Sanjaye Ramgoolam and Congjun Wu for email correspondences. This work was partially supported by a Grant-in-Aid for Young Scientists (B) (Grant No.23740212), Overseas   Dispatching Program 2013 of National College of Technology, and The Emma Project for Art and Culture.   


\begin{thebibliography}{99}



\bibitem{RyuSFL2009}
Shinsei Ryu, Andreas Schnyder, Akira Furusaki, Andreas Ludwig
{\it``Topological insulators and superconductors: ten-fold way and dimensional hierarchy''},  
 New J. Phys. 12 (2010) 065010;  arXiv:0912.2157. 
\bibitem{Hosuretal2009}
Pavan Hosur, Shinsei Ryu, Ashvin Vishwanath, {\it``Chiral Topological Insulators, Superconductors and other competing orders in three dimensions''}, Phys. Rev. B, 81 (2010) 045120; arXiv:0908.2691. 
\bibitem{NeupertSRChM2012}
Titus Neupert, Luiz Santos, Shinsei Ryu, Claudio Chamon, Christopher Mudry,    
{\it``Noncommutative geometry for three-dimensional topological insulators''}, 
Phys. Rev. B 86 (2012) 035125;   arXiv:1202.5188.
\bibitem{WngDD2014}
Sheng-Tao Wang, Dong-Ling Deng, Lu-Ming Duan, {\it``Probe of Three-Dimensional Chiral Topological Insulators in an Optical Lattice''}, arXiv:1402.1204. 
\bibitem{EstienneRB2012}
B. Estienne, N. Regnault, B. A. Bernevig,    
{\it``D-Algebra Structure of Topological Insulators''}, 
Phys. Rev. B 86 (2012) 241104(R);  arXiv:1202.5543.
\bibitem{Nambu1973}
Yoichiro Nambu,  
{\it ``Generalized Hamiltonian Dynamics''}, Phys.Rev.D7 (1973) 2405-2412.  
\bibitem{CurtrightZachos2003}
Thomas Curtright, Cosmas Zachos, 
{\it ``Classical and Quantum Nambu Mechanics''}, Phys.Rev.D68 (2003) 085001; hep-th/0212267.  
\bibitem{DeBellisSS2010}
Joshua DeBellis, Christian Saemann, Richard J. Szabo,
{\it ``Quantized Nambu-Poisson Manifolds and n-Lie Algebras''}, J.Math.Phys.51 (2010) 122303; arXiv:1001.3275. 
\bibitem{BasuHarvey2004}
Anirban Basu, Jeffrey A. Harvey, 
{\it``The M2-M5 Brane System and a Generalized Nahm's Equation''}, 
Nucl.Phys. B713 (2005) 136-150; hep-th/0412310. 
\bibitem{BaggerLambert2006}
Jonathan Bagger, Neil Lambert, 
{\it``Modeling Multiple M2's''}, 
Phys.Rev.D75 (2007) 045020; hep-th/0611108. 
\bibitem{Gustavsson2007}
Andreas Gustavsson, 
{\it``Algebraic structures on parallel M2-branes''}, 
 Nucl.Phys.B811 (2009) 66-76; arXiv:0709.1260. 
\bibitem{BaggerLambert2007}
Jonathan Bagger, Neil Lambert, 
{\it``Gauge Symmetry and Supersymmetry of Multiple M2-Branes''}, 
Phys.Rev.D77 (2008) 065008; arXiv:0711.0955. 
\bibitem{Hasebe2014-1}
Kazuki Hasebe,    
{\it``Higher dimensional quantum Hall effect as A-class topological insulator''}, arXiv:1403.5066. 
\bibitem{Nair-Daemi-2004}
V.P. Nair, S. Randjbar-Daemi,    
{\it``Quantum Hall effect on $S^3$, edge states and fuzzy $S^3/{\bf Z}_2$''}, 
Nucl.Phys. B679 (2004) 447-463; hep-th/0309212.
\bibitem{Haldane1983} 
F.D.M. Haldane,
{\it``Fractional quantization of the Hall effect: a hierarchy of incompressible 
quantum fluid  states"}, 
 Phys. Rev. Lett. 51 (1983) 605-608. 
\bibitem{ZhangHu2001} 
 S.-C. Zhang, J.-P. Hu,
{\it``A four-dimensional generalization of the quantum Hall effect"}, 
Science 294 (2001), no. 5543, 823-828; cond-mat/0110572. 
\bibitem{Ramgoolam2002}
Sanjaye Ramgoolam,    
{\it``Higher dimensional geometries related to fuzzy odd-dimensional spheres''}, 
JHEP 0210 (2002) 064; hep-th/0207111.
\bibitem{JabbariTorabian2005}
M. M. Sheikh-Jabbari, M. Torabian,  
{\it``Classification of All 1/2 BPS Solutions of the Tiny Graviton Matrix Theory''}, 
JHEP 0504 (2005) 001; hep-th/0501001. 
\bibitem{Halperin1983}
B. I. Halperin,  
{\it``Theory of the quantized Hall conductance''}, 
Helv. Phys. Acta  56 (1983) 75. 
\bibitem{Yoshiokaetal1988}
D. Yoshioka, A. H. MacDonald, S. M. Girvin,  
{\it``Connection between spin-singlet and hierarchical wave functions in the fractional quantum Hall effect''},  
Phys. Rev. B 38 (1988) 3636-3639. 
\bibitem{WenZee1992}
X.G. Wen, A. Zee,  
{\it``Classification of Abelian quantum Hall states and matrix formulation of topological fluids''},  
Phys. Rev. B 46 (1992) 2290-2301. 
\bibitem{BernevigZhang2006}
B.A. Bernevig and S.C. Zhang,, 
{\it``Quantum Spin Hall Effect''},  Phys. Rev. Lett. 96,
106802 (2006); cond-mat/0504147. 
\bibitem{LiWu2013}
Yi Li, Congjun Wu, 
{\it``High-Dimensional Topological Insulators with Quaternionic Analytic Landau Levels''}, 
Phys. Rev. Lett. 110 (2013) 216802; arXiv:1103.5422. 
\bibitem{Li-I-Y-W-2012}
Yi Li, Kenneth Intriligator, Yue Yu, Congjun Wu, 
{\it``Isotropic Landau levels of Dirac fermions in high dimensions''},  
Phys. Rev. B  85 (2012) 085132; arXiv:1108.5650. 


\end{thebibliography}
\end{document}